\documentclass[10pt]{wlscirep}
\title{Limits of Risk Predictability in a Cascading Alternating Renewal Process Model}

\usepackage{url}
\usepackage{subcaption}
\usepackage{caption}
\usepackage{tabularx}
\usepackage{array}
\usepackage{epstopdf}

\author[1,2]{Xin Lin}
\author[1,3]{Alaa Moussawi}
\author[1,3]{Gyorgy Korniss}
\author[4]{Jonathan Z. Bakdash}
\author[1,2,*]{Boleslaw K. Szymanski}

\affil[1]{Social and Cognitive Networks Academic Research Center, Rensselaer Polytechnic Institute, Troy NY, 12180, USA}
\affil[2]{Dept. of Computer Science, Rensselaer Polytechnic Institute, 110 8th Street, Troy NY, 12180, USA}
\affil[3]{Dept. of Physics, Applied Physics and Astronomy, RPI, 110 8th Street, Troy NY, 12180, USA}
\affil[4]{U.S. Army Research Laboratory, Aberdeen Proving Ground MD, 21005, USA}

\affil[*]{szymab@rpi.edu}

\begin{abstract} 
Most risk analysis models systematically underestimate the probability and impact of catastrophic events (e.g., economic crises, natural disasters, and terrorism) by not taking into account interconnectivity and interdependence of risks. To address this weakness, we propose the Cascading Alternating Renewal Process (CARP) to forecast interconnected global risks. However, assessments of the model’s prediction precision are limited by lack of sufficient ground truth data. Here, we establish prediction precision as a function of input data size by using alternative long ground truth data generated by simulations of the CARP model with known parameters. We illustrate the approach on a model of fires in artificial cities assembled from basic city blocks with diverse housing. The results confirm that parameter recovery variance exhibits power law decay as a function of the length of available ground truth data. Using CARP, we also demonstrate estimation using a disparate dataset that also has dependencies: real-world prediction precision for the global risk model based on the World Economic Forum Global Risk Report. We conclude that the CARP model is an efficient method for predicting catastrophic cascading events with potential applications to emerging local and global interconnected risks.
\end{abstract}

\begin{document}
\flushbottom
\maketitle

\section*{Introduction}
A generalized Alternating Renewal Process model referred to as Cascading Alternating Renewal Process (CARP), has been recently proposed for dynamically modeling a global risk network represented as a set of Poisson processes \cite{Szymanski2015}. The aim was the recovery of the directly observable and hidden parameters of the model using historical data to predict future activation of global risks, and especially cascades of such spreading risk activations \cite{Watts2002, MotterLai2002}. This approach raises questions about the reliability of such recovery and the dependence of the prediction precision on the complexity of the model and the length of its historical data. Since most of the world's critical infrastructures, including the global economy, form a complex network that is prone to cascading failures \cite{wef15} this question is important but difficult to answer given the for lack of ground truth data. 

Here, we discuss how the limits of prediction precision can be established for a specific system as a function of the input data size by simulating the CARP model with known parameters to generate many alternative ground truth datasets of arbitrary length. This enables us to the given system to establish the precision bound with the given length of historical data. We illustrate our approach with a model of fire risks in an artificial city that consists of modular blocks of diverse housing that are easy to assemble into an ever-growing complex system. We first simulate the fires in such cities of varying sizes, over varying time periods, in each case using different random generator seeds to create many alternative ground truth datasets. Then, using Maximum Likelihood Estimation (MLE) \cite{Dempster1977}, we recover the parameters and compare them with their values used for simulations. We also measure the recovered parameter precision as a function of (i) the complexity of the system (in our case the size of the cities) and (ii) the number of events present in the historical data. Finally, using real-world data with the developed methodology, we assess the precision of parameter recovery in the World Economic Forum model \cite{Szymanski2015}.

\subsection*{Risk Modeling}
Most quantitative risk analysis models (e.g., Value at Risk and Probability-Impact models) systematically underestimate the probability and impact of worst-case scenarios (i.e., maximum loss for a given confidence level, typically "tail" outcomes) for catastrophic events such as economic crashes, natural disasters, and terrorist attacks \cite{banks2005catastrophic}. Underestimation in such models is due to speciously assuming the sequences of random variables in probability distributions are normal, independent, and identically distributed (IID) \cite{taleb2007black, pate2012black} thus discounting the potential of interdependencies and interconnections between events.  

Few quantitative risk models capture the non-IID properties of risk factors and their impacts \cite{national2010}. The Havlin model uses branching to predict cascading failures (e.g., power grids, communication networks) \cite{buldyrev2010catastrophic}. A small fraction of initial failures could cause catastrophic damage in mutually dependent systems. The authors used the percolation theory and detected a phase transition for the robustness and functionality of the interdependent networks. The Ganin model of resilience \footnote{Resilience has a variety of definitions, but the common definition is the planning, preparation, absorption, recovery, and adaptation of systems under negative events \cite{Resilience2016}. Predicting the limits of risk is consistent with the common definition.} provides an analytical definition to criticality and a method for determining it to design more resilient technical systems  \cite{ganin2016}. While both techniques go beyond simple risk models, which assume risks exhibit independent probabilities and impacts, neither aims to quantify the limits of predictability for interconnected risks. In contrast, the CARP model is a novel method in which interdependencies and interconnections are explicitly represented, and the model parameters are recovered from historical data using Maximum Likelihood Estimation. Moreover, the CARP model offers a quantitative risk assessment for interconnected conceptual models such as Reason's Swiss cheese model of failure \cite{Reason1900}. In Reason's model, defenses preventing failure are layers (of Swiss cheese) and when the holes of the layers align a risk may materialize. Previously, Reason's model has been formalized using percolation theory \cite{Frosch2006}, but this formalization has not been validated.

\subsection*{Alternating Renewal Process}
The definition of Alternating Renewal Process (ARP) originated in renewal probability theory. A simple renewal process alternates between two states: the normal state, which accounts for its operational time and the abnormal (failure or repair) state, which accounts for its holding (non-operational) time. Each state is governed by a Poisson process specific to this state \cite{Cinlar1969, Cox1965}. The ARP can be used to find the best strategy for replacing worn-out machinery \cite{Dickey1990, Weide2015}.

The CARP model expands ARP in several ways. The most important extension is to allow for defining risks as a network of risks with the given set of weighted edges representing interdependencies and probabilities of state transitions associated with each edge. Since state transitions are often observed without distinction for the inner or external, edge induced transitions, both probabilities are treated as hidden variables, since only their joined effect is observed directly. Another generalization is not restricting the number of states to two. Finally, we associate with CARP an MLE procedure for recovery of CARP parameters from historical data of the system evolution over time. 

\subsection*{Model Structures}
The city modeled is diverse. A small fraction of houses is large and placed sparsely with low fire propagation probabilities and high recovery rates. A larger fraction of houses with a medium size have a higher spatial density and fire propagation probabilities. Finally, the most densely packed small houses have the highest fire propagation probabilities, and lowest recovery rates. 

\begin{figure*}[!ht]
\begin{center}
\includegraphics[width = 0.8\textwidth,height = 4.0 in]{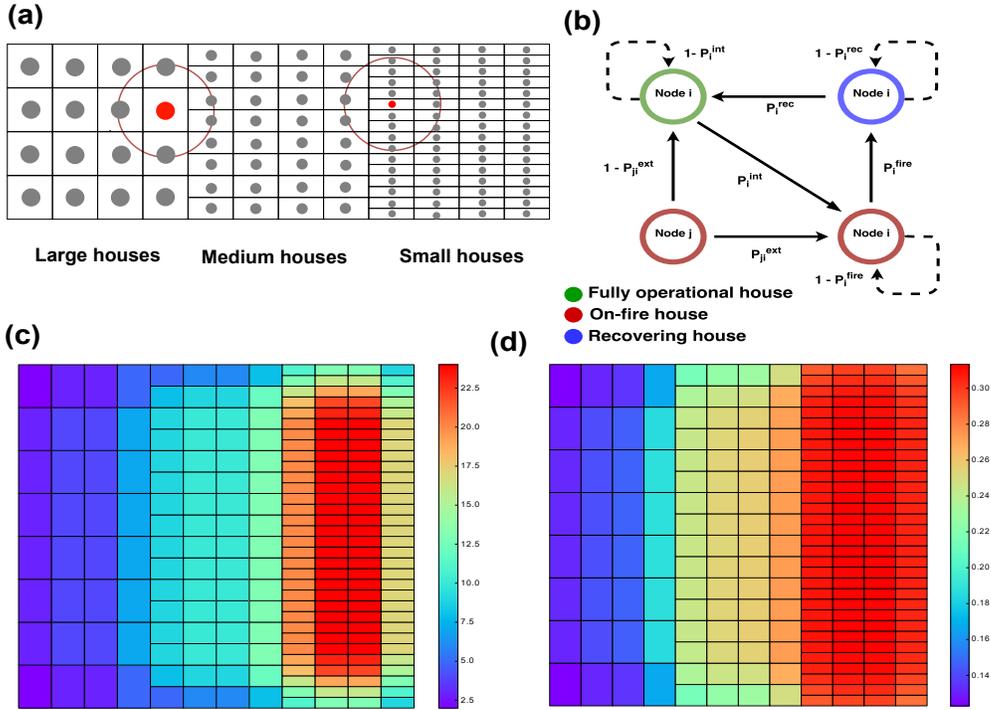}
\end{center}
\caption{ {\bf (a) Basic block of a city.} A sample city consists of four basic blocks. Three types of houses are represented as nodes. Red circles show the range of external fire spread. All houses in this circle may catch fire from the burning center node. {\bf (b) Fire propagation dynamics.} This diagram shows state transitions. The nodes represent a fully operational house (green), a burning house (red) and a recovering house (blue). The lines show state transitions to a new state as solid lines, and state transitions to the same state as dotted lines. {\bf (c) House degrees in a $8$ block $224$ house city.} The degree increases horizontally from left to right as house types change from large to medium and to small. Centrally placed small houses have the highest degrees. {\bf (d) Fraction of time each house is on fire.} Results are averaged over $100$ independent realizations. Each simulation runs $10^6$ time units. The houses are fully operational initially. The simulations used parameter values listed in Table \ref{tab:Poisson intensity}.}
\label{fig:figure_1}
\end{figure*}

Houses are grouped into blocks stitched together into a continuous city as shown in Fig. \ref{fig:figure_1}(a). Large and small houses sit on the West and East boundaries of the city, respectively. The North and South boundaries border all three types of houses. The houses are placed on rectangular lots whose size is commensurate with the size of the house. A large house occupies a square lot of one by one unit size. There are four such squares horizontally laid out with no space in between. Each medium house has a rectangular lot of half unit length vertically by one unit length horizontally. Each block holds $8$ medium houses. Finally, repeating a similar pattern, small houses sit on a rectangular lot of one by a quarter unit size. One block contains $16$ small houses. Distances between two adjacent houses vary from $1$ for large houses to $ 1/2$ for medium and $1/4$ for small houses. A basic block holds $28$ houses on $12$ square units of land. A city of arbitrary large size can be built by repeatedly adding blocks to it. Fig. \ref{fig:figure_1}(c) shows the degree of each house, which is defined as the number of houses within a fixed distance. As shown in Fig. \ref{fig:figure_1}(a), all houses within a fixed distance may catch fire from the burning center house. Hence, the small houses have a relative higher degree because of the high density of nearby houses. The surrounding houses on the boundary of the city and the border of different communities experience a slightly smaller degree compared with other same-size houses. As we explain later, we assume that, the likelihood of a house catching fire is determined by two factors: the materials the house is made of and the density of its neighbors.

\noindent
\section*{Results}
Unlike real-life, in simulations, we can arbitrarily vary the length of time over which we collect historical data and produce many variants of such data to measure the prediction precision of our model. We start with a mixed model with $8$ large houses, $16$ medium houses and $32$ small houses, for a total of $56$ houses. The parameter recovery is applied at seven different intervals: $100$, $200$, $400$, $800$, $1600$, $3200$ and $6400$ time steps. The parameter recover is run on $50$ versions of historical data created by simulations running with different seeds for the random number generator to account for the randomness of the stochastic processes. To quantify the accuracy of parameter recovery, we compute the relative error between the recovered values and the target values used for creating ground truth data. 

\begin{figure*}[!ht]
\begin{center}
\includegraphics[height = 3.0 in]{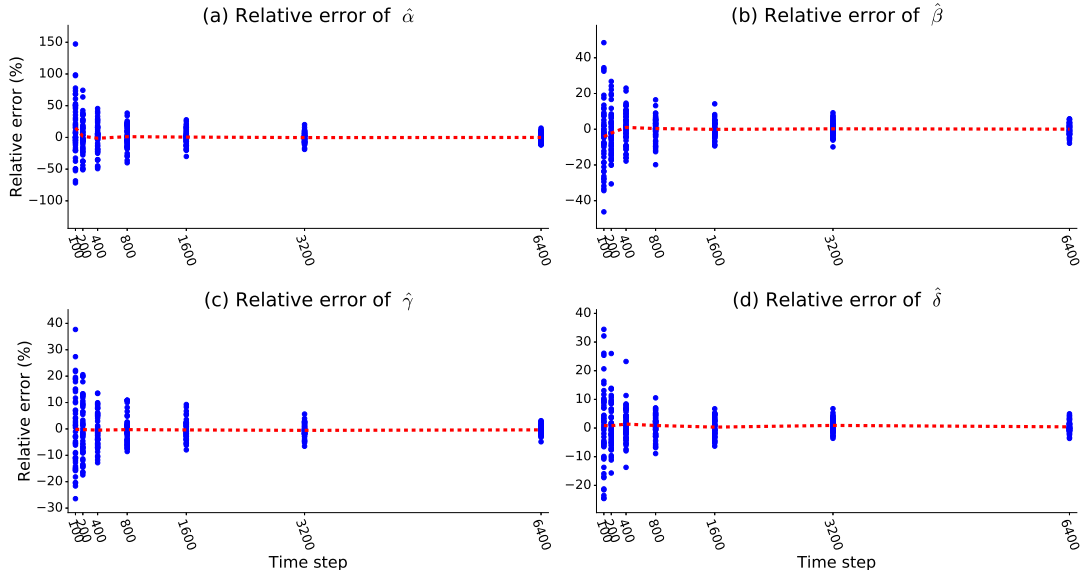}
\end{center}
\caption[parameter recovery in fire-propagation model]
{{\bf Parameter recovery in the fire-propagation model.} The x-axis includes seven time intervals: $100$, $200$, $400$, $800$, $1600$, $3200$ and $6400$ time steps. The y axis shows the relative error for the recovered parameters. Using parameter values shown in Table \ref{tab:Poisson intensity}, parameter recovery was run on $50$ different historical datasets generated by simulations with different seeds for the random number generator; the results of these runs are represented by blue dots. The red dashed curves show the average values of the relative error. The visible trend is that the average of relative errors tends asymptotically to zero and the variance exhibits power law decrease as the number of time steps increases, which means more training data improves the performance of parameter recovery.} 
\label{fig:parameter recovery in fire-propagation model}
\end{figure*}

As shown in Fig. \ref{fig:parameter recovery in fire-propagation model}, the blue vertical dots represent the relative error of parameter recovery in one realization. There are $50$ blue dots for each particular length of the training dataset. The scattering of blues dots represents the variation of parameter recovery accuracy. The red curve is the average value among these $50$ realizations, which is very close to zero. Obviously, the variance is a better measurement for the precision since variance considers the positive and negative errors instead of canceling them out. The average of relative errors tends asymptotically to zero and the variance shrinks according to the power law with an increase in sample size which is the length of historical data series in this case. This trend is consistent with the asymptotic behavior of the MLE method \cite{Cramer1946}. When the sample size is very large, the relative error of realization follows a normal distribution with $0$ mean value and a finite yet small variance. More data decreases relative error, and therefore it is useful to find a balance between run time and prediction accuracy. 

The relative errors of $\hat{\alpha}$ and $\hat{\beta}$ are larger than that of other parameters. This is due to the combined effects of two Poisson processes causing the same transition. As defined, $\alpha$ and $\beta$ represent the intensity of internal and external fire ignition processes respectively. In real life, it is often hard to determine the actual reason. During the parameter recovery, these two parameters influence chances of each other to start a fire, which impacts the computation of likelihood function. This nonlinear effect tangles the errors of $\hat{\alpha}$ and $\hat{\beta}$ together as larger value of $\hat{\alpha}$ can be compensated by smaller value of $\hat{\beta}$ and vice versa.       

\begin{figure*}[!ht]
\begin{center}
\includegraphics[height = 2.2 in]{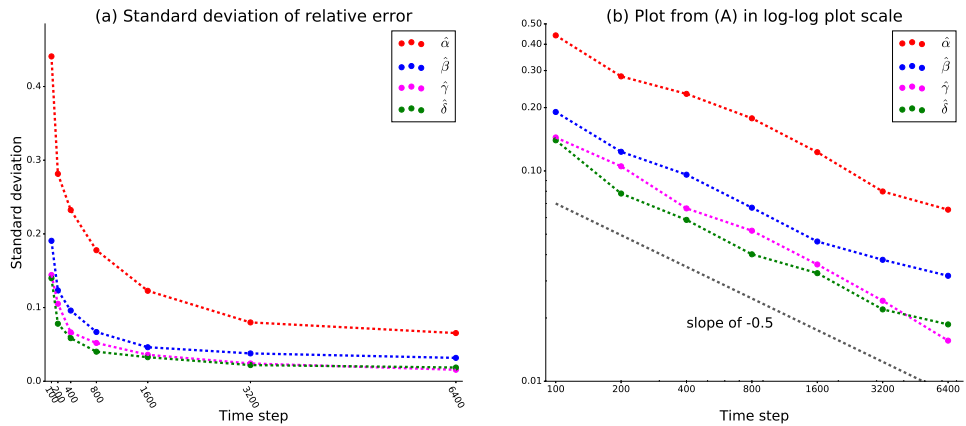}
\end{center}
\caption {{\bf Comparison of standard deviation of relative error of parameter recovery.} We use seven simulation time intervals: $100$, $200$, $400$, $800$, $1600$, $3200$ and $6400$ time steps. There are $56$ houses. Using parameter values shown in Table \ref{tab:Poisson intensity}, parameter recovery was run on 50 different historical datasets generated by simulations with different seeds for random number generator. The plots show standard deviation of the relative error of recovered parameters averaged over $50$ runs in linear scale (a) and logarithmic scale (b).} 
\label{fig:std plots}
\end{figure*}

We also compute the standard deviation of relative error of recovered parameters. Fig. \ref{fig:std plots}(a) shows that the distribution of the standard deviation follows a power law. The standard deviation decreases very quickly and then slowly as historical data size increases. The double-logarithmic plot in Fig. \ref{fig:std plots}bB) has slope close to $-0.5$, which shows that the standard deviation decreases in a power-law fashion as the training data size increases. 

For real-world case processes, it is impossible to get multiple historical datasets for parameter recovery, yet recovery error based on single dataset may be different from the one based on the average of such recovered values on multiple datasets. Hence, we study how sensitive our methodology is to imperfect input datasets. In order to test this sensitivity, we compare four cases of simulations in which we record the average length of time in each state and the number of emerging fires during four simulation periods: $400, 800, 3200$ and $6400$. The first case is using the target values of parameters. The second case is using the averaged value of parameters recovered from $50$ independent realizations. The third case employs adding one standard deviation $\sigma$ to the average value of recovered parameters. The last case is subtracting one standard deviation $\sigma$ from the average value of recovered parameters. The four sets of parameters employed in our simulation are listed in Table \ref{tab:simulation comparison}:

\begin{table}[ht]
\centering
\begin{tabular}{|c|c|c|c|c|}
\hline
Parameters & $\alpha$ & $\beta$ & $\gamma$ & $\delta$ \\
\hline
Target value & 0.00800 & 0.01200 & 0.01600 & 0.03200\\
\hline
Recovered value & 0.00799 & 0.01199 & 0.01596 & 0.03204 \\
\hline
Recovered value $+ \, \sigma$ & 0.00853 & 0.01162 & 0.01622 & 0.03274 \\
\hline
Recovered value $- \, \sigma$ & 0.00747 & 0.01237 & 0.01570 & 0.03134 \\
\hline
\end{tabular}
\caption{\label{tab:simulation comparison}Parameter values for simulation of historical date with range of parameter values.}
\end{table}

\begin{figure*}[!ht]
\begin{center}
\includegraphics[height = 3.5 in]{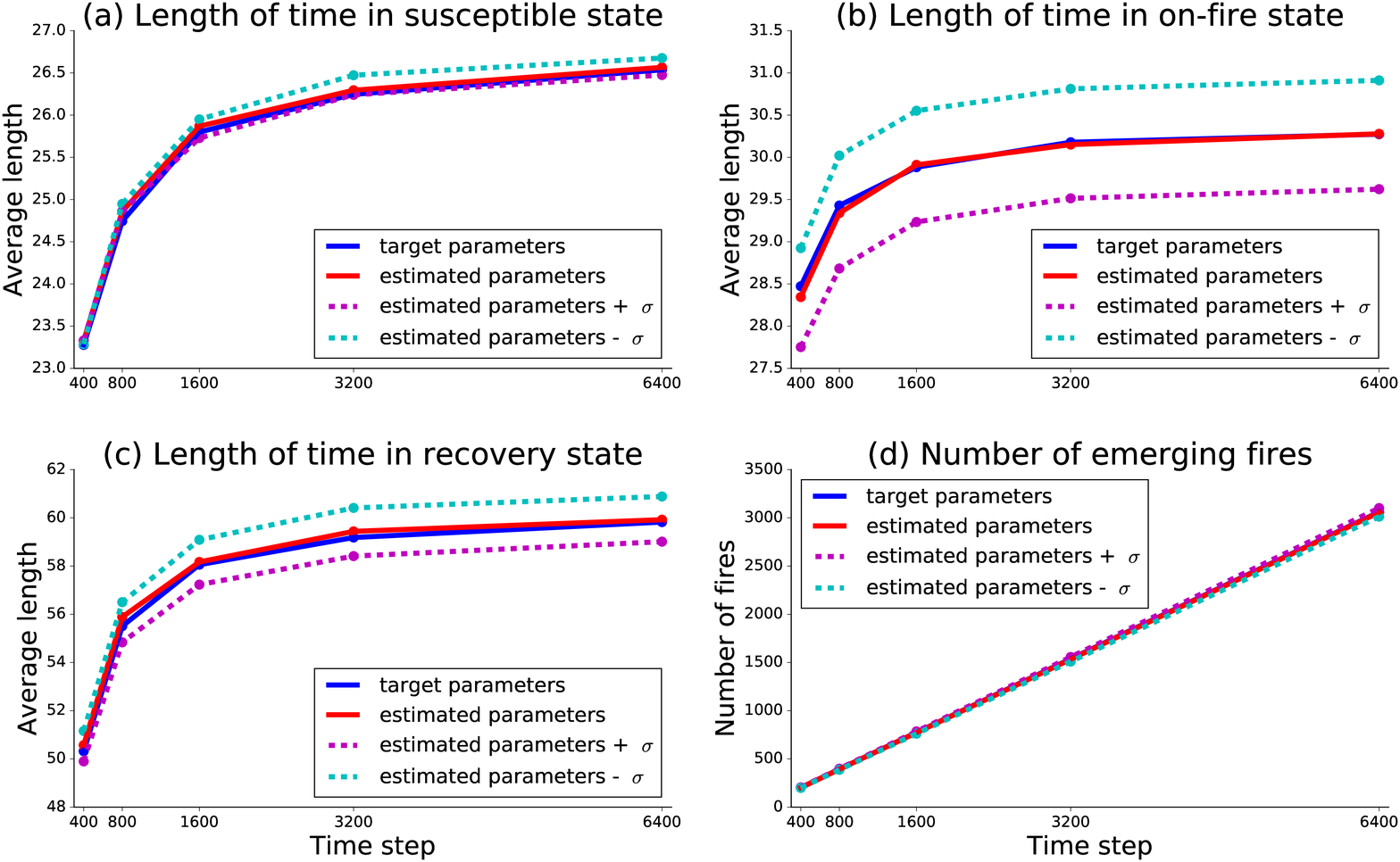}
\end{center}
\caption {{\bf Impact of recovered parameter uncertainty on the prediction of model dynamics.} The average length of time in each state is recorded for five time intervals: $400$, $800$, $1600$, $3200$ and $6400$ time steps. The number of houses is $56$. Using parameter values shown in Table \ref{tab:simulation comparison}, parameter recovery was run on $50$ different historical datasets generated by simulations with different seeds for random number generator. (a) shows how long a house stays in the fully operational state which is determined by internal and external fire triggering processes. Since the effect of increasing $\alpha$ is reduced by the effect of decreasing $\beta$ and vice versa, their combined effects is smaller than in the case of other parameters. (b-c) show the average length of time for the states of on-fire and recovery. Only one parameter determines the length of time so the gap is larger than (a) but still low. (d) shows the number of new fires during the simulation. All cases yield similar results.} 
\label{fig:simulation comparison}
\end{figure*}

The average values of recovered parameters come from $50$ independent realizations with $6400$ time steps of training data. The standard deviations are: $\sigma_{\alpha} = 0.00053, \sigma_{\beta} = 0.000372, \sigma_{\gamma} = 0.00026, \sigma_{\delta} = 0.0007$. When adding one standard deviation $\sigma$ to average value of $\hat{\alpha}$, we subtract $\sigma$ from $\hat{\beta}$ and vice versa. We assume complementary recovery errors on $\alpha$ and $\beta$ since both of them contribute to the fire triggering process so the maximum value of one is likely to be reached at the minimum of the other. This corresponds to the assumption that the unique historical dataset produces parameter values within one $\sigma$ of their average values, more stringent assumption might require using broader interval around average values of parameters. In simulation, we record how long each house stays fully operational, on-fire and in recovery and record the number of new fires during the simulation. We generate over $50$ different historical datasets and gather results upon reaching five simulation periods: $400, 800, 1600, 3200$ and $6400$. With each dataset, we run $50$ independent realizations to assess the predictability of our methodology. There is a trade-off between the running time and precision. To save computational effort, we set the number of independent realizations at a moderate value of $50$. Fig. \ref{fig:simulation comparison} shows that all four parameter estimations have similar precision. In Fig. \ref{fig:simulation comparison}(a), the average duration of the fully operational state is almost the same for four cases since two parameters $\alpha$ and $\beta$ have opposite influences on fully operational houses. In Fig. \ref{fig:simulation comparison}(b-c), the gap between the cases of estimated parameters $\pm \sigma$ is very small. In Fig. \ref{fig:simulation comparison}(d), the number of emerging fires for all cases are increasing linearly as a function of time. The largest relative error is in predicting the length of fire, and even that is just below $2\%$. For simulation efficiency,  we set the target values of parameters in such a way that the duration of staying in a state is short and of the same order for all states, while in reality they are different, for example for any house staying on fire is orders of magnitude shorter than getting rebuilt. We intentionally set large and comparable target values to generate sufficient state transitions and test the quality of parameter recovery in an arbitrary case. Table \ref{tab:simulation comparison} shows that the estimated parameters approach the target values very well, even if the target values are of the same order as the initial settings.

\begin{figure*}[!ht]
\begin{center}
\includegraphics[height = 3.5 in]{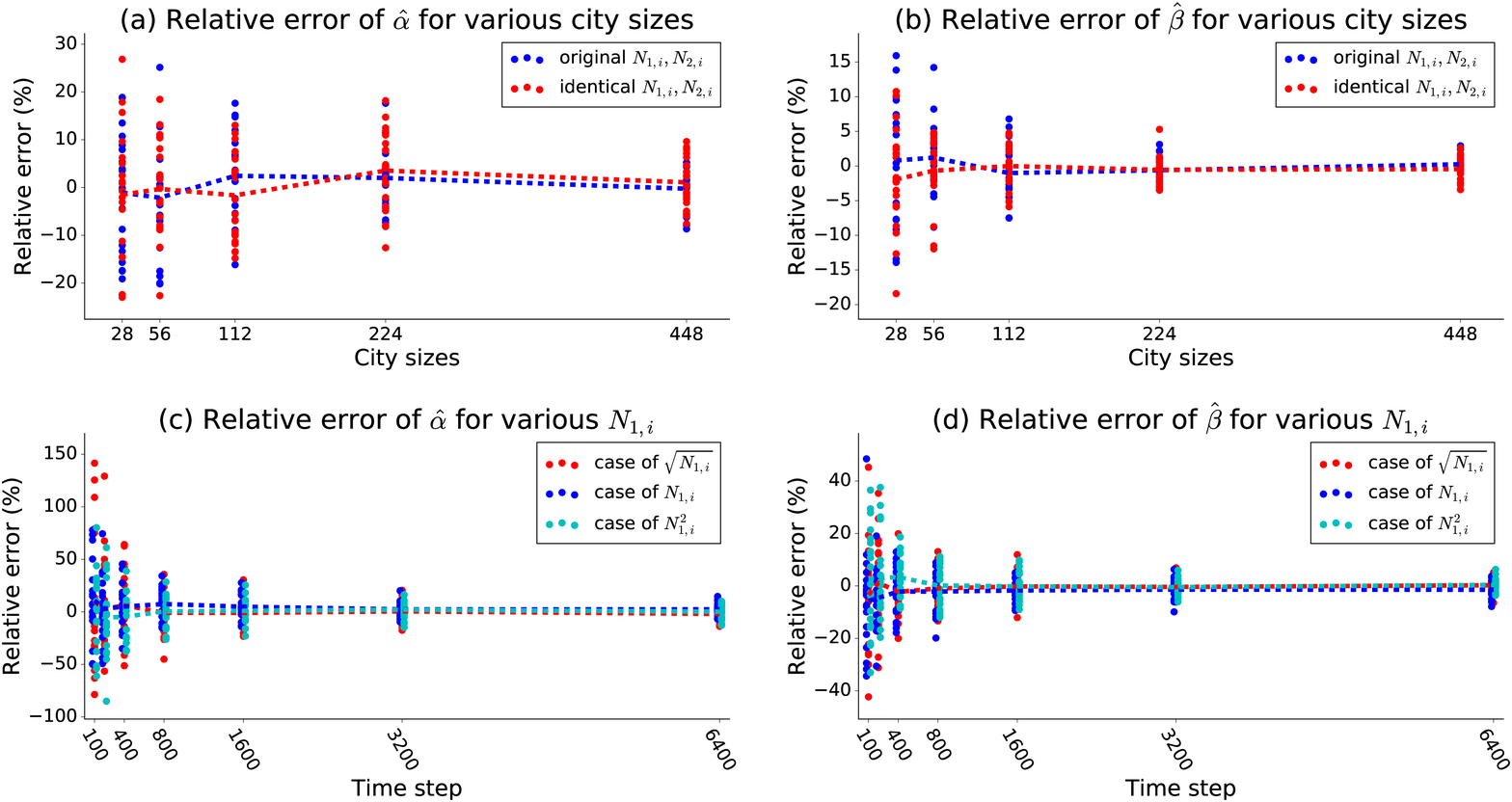}
\end{center}
\caption {{\bf Parameter recovery in various scenarios.} (a-b) show relative errors of recovered $\hat{\alpha}$ and $\hat{\beta}$ for five city sizes: $28$, $56$, $112$, $224$ and $448$. Blue color represents the city with three types of houses and different values of $N_{1,i}$ and $N_{2,i}$. Red color represents the city with identical value of $N_{1,i}$ and $N_{2,i}$. The dashed curve shows the mean values of relative error over $20$ realizations and each dot represents one realization. The length of historical dataset is $1600$. Parameter values are shown in Table \ref{tab:simulation comparison}. (c-d) show relative errors of recovered $\hat{\alpha}$ and $\hat{\beta}$ for three cases of $N_{1,i}$. Red dots represent the case of $\sqrt{N_{1,i}}$, blue represents the case of $N_{1,i}$, and cyan represents the case of $N_{1,i}^2$. The results come from $20$ independent realizations and $6$ different lengths of historical dataset: $100$, $400$, $800$, $1600$, $3200$ and $6400$. $N_{1,i}$ is $0.4$ for large $0.3$ for medium and $0.2$ for small houses. $\alpha$ = $0.08$, $\beta$ = $0.012$, $\gamma$ = $0.016$ and $\delta$ = $0.032$.} 
\label{fig:recovery precision}
\end{figure*}

In addition to the parameter recovery from different lengths of historical datasets, we study how the precision of parameter recovery varies against the complexity of the system and the standard deviation of number of fires starting each day. Fig. \ref{fig:recovery precision}(a-b) show the relative errors of recovered parameters for various city sizes (number of houses): $28$, $56$, $112$, $224$ and $448$. Here, we compare two cases. In the first case, we keep three types of houses and use the parameter values shown in Table \ref{tab:simulation comparison}. The blue dots in Fig. \ref{fig:recovery precision}(a-b) represent the results of the first case. In the second case, we initialize all houses with the same value of $N_{1,i} = 0.3$ and $N_{2,i} = 0.3$, which are equal to the values for middle houses. The red dots in Fig. \ref{fig:recovery precision}(a-b) represent the results from this case. Therefore, we remove the influence of house types on the results. In both cases, the length of historical data is $1600$ time steps and we run over $20$ different datasets. Only two parameters ($\hat{\alpha}$ and $\hat{\beta}$) are shown in this figure since they are involved with the fire igniting process and another two parameters have similar trend. We find that the parameter recovery in a larger city has a smaller mean value and variance of relative errors. As the city's size increases, there are more state transitions within specific periods, leading to a more precise recovery of our parameters. To study the impact of intensity of emerging fires on recovery precision, we change the values of $N_{1,i}$, which define the intensities of fires. We compare three cases: $\sqrt{N_{1,i}}$ in red, $N_{1,i}$ in blue and $N_{1,i}^2$ in cyan. Fig. \ref{fig:recovery precision}(c-d) show the relative error of recovered $\hat{\alpha}$ and $\hat{\beta}$. The model with more emerging fires at each day enjoys a smaller variance of relative errors. The reason is similar to that shown above: more state transitions in the historical dataset and higher precision of the recovered parameters.       

\subsection*{Parameter Recovery Precision in Global Risk Network}
Here, we show how to apply the presented approach to a disparate, real-life dataset of the global risk network, which, as with fires in cities, exhibits spreading risk activation \cite{Szymanski2015}. Using CARP, we estimate hidden parameters of global risks previously modeled using an Alternating Renewal Process \cite{Szymanski2015}. Experts from the World Economic Forum $2013$ Global Risks Report \cite{wef13} define the properties of $50$ global risks grouped into five categories: economic, environmental, geopolitical, societal, and technological. These assessments include the likelihood, impact of materialization, and connections of each risk. We take the advantage of this crowd-sourcing assessment to build an interconnected network to simulate risk propagation through the system. 

In the global risk network, each risk has binary states (normal and active), and the state transitions follow Poisson processes. The difference is that in the global risk network, there are three state transitions instead of four in the fire-propagation model: $1$, internally igniting fire process transitioning the risk from normal to active state; $2$, externally starting fire process also transitioning risk from normal to active state; and $3$, starting recovery process transitioning risk from active to normal state. Hence, we have three control parameters ($\alpha, \beta, \gamma$) and recover the optimal parameter values from historical events. Using data collected over a $156$-month period and Maximum Likelihood Estimation, we recover the following parameter values: $\alpha = 0.003038$, $\beta = 0.00117$ and $\gamma = 3.5561$. They control internal risk materialization ($\alpha$), external risk materialization ($\beta$) and recovery ($\gamma$) processes, and their detailed definitions can be found in Ref. \cite{Szymanski2015}. Once we obtain the parameter values, we can simulate the stochastic processes driving model evolution. As in the case of the fire-propagation model presented here, we use these recovered parameters as the ground truth parameters for establishing the parameter recovery precision.

We apply the same method used in Fig. \ref{fig:simulation comparison} to this global risk network. For the precision evaluation, we obtain multiple recovered-parameters from different lengths of alternative historical data to predict the number of risk materialization for a selected period repeatedly with different random generator seeds. First, based on the ground truth parameters recovered from the real historical data ($\alpha, \beta, \gamma$), we generate $120$, $240$, $480$ and $960$ monthly time steps of alternative historical data (representing $10$, $20$, $40$ and $80$ years) for parameter recovery. Then, $125$ cases of parameter recovery ($\alpha_i, \beta_i, \gamma_i, i = 1...125$) are finished for each period of time. Next, using the recovered parameters from both real and alternative historical data, we complete $20$ realizations for a prediction of $4$ periods: $120$, $240$, $480$ and $960$. In each period, we use the average value of risk materialization among all $20$ realizations to measure the performance of each simulated case. After that, we compute the relative error of the average number of risk materializations between the case with alternative recovered-parameters ($\alpha_i, \beta_i, \gamma_i, i = 1...125$) and the ground truth parameters ($\alpha, \beta, \gamma$). In the end, we remove the $39$ simulation cases with the worst performance (with the largest absolute relative error). The remaining $86$ cases ($68.8\%$ of $125$ cases) of results determine the $\pm \sigma$ boundary for the performance of the system with estimated parameters.  

\begin{figure*}[!ht]
\begin{center}
\includegraphics[height = 3.5 in]{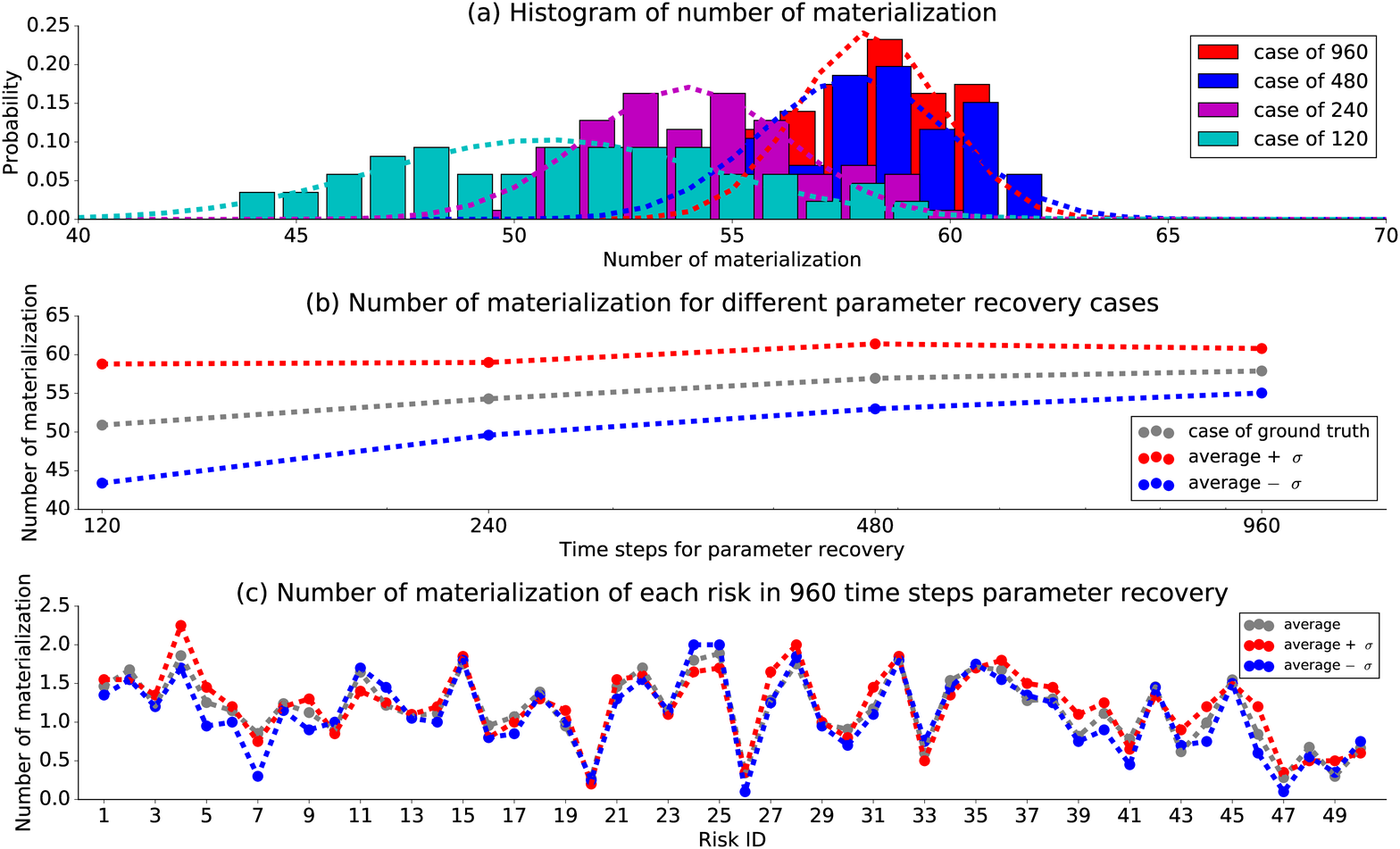}
\end{center}
\caption {{\bf Performance comparison of recovered parameters in the global risk network.}  There are $125$ cases of estimated parameters. $4$ different periods of time steps ($120$, $240$, $480$ and $960$) are used to estimate parameters and simulate future behaviors. For each set of estimated parameter, we finish $20$ realizations and average the number of materialization during the simulation. $\pm\sigma$ boundary is determined by removing $39$ sets of estimated parameters with worst performance from ground truth case. (a) Histogram of number of materialization. (b) shows the boundary of $\pm \sigma$ performance for number of materialization in each period. (c) shows the number of materialization for each risk in the case of $960$ time steps.}
\label{fig:parameter recovery in wef}
\end{figure*}

Fig. \ref{fig:parameter recovery in wef}(a) shows a histogram of the number of materializations in each case. Dash curve represents a Gaussian fitting over the histogram. As the length of simulation period increases, the distribution of the number of materialization gradually approaches a normal distribution. Meanwhile, the distribution shifts to the right and gets close to a steady level. In the case with longer time steps for parameter recovery and prediction simulation, the predictability is more consistent, and variance shrinks faster. Fig. \ref{fig:parameter recovery in wef}(b) shows the boundary of $\pm \sigma$ performance for the number of materializations in each period. It is evident that the distance between the boundaries is decreasing as we increase the length of the period, which implies a higher precision of prediction. Fig. \ref{fig:parameter recovery in wef}(c) shows the number of materializations for each risk in the case of $960$ time steps. The difference between three cases is slight indicating a consistent prediction of estimated parameters.

\section*{Discussion}
The CARP model is used to simulate and then recover parameters of heterogeneous stochastic processes. First, we created a model of fires in the cities that we use to illustrate our approach. Using assumed parameters values, we generated several historical datasets and used them to measure parameter recovery precision. The results confirmed that the accuracy of our method increases as the amount of data increases even in the presence of parameters hidden from direct observations.

Applying our approach to real-life cases, we started with the recovery of the model parameter values based on unique and limited real-life ground truth data. Then, using these values as ground truth, we finished simulations to create many alternative historical datasets. Then using these historical datasets, the parameters are recovered by applying MLE method. Next, we compared the results to the assumed ground truth values to measure the accuracy of recovery. The standard deviation of relative error of parameter recovery exhibits a power-law decay with an exponent value of $-0.5$ as the training data size increases. The resulting statistics enable us to verify the reliability of predictions based on originally recovered parameters. We did so by comparing original predictions to predictions based on parameters differing from their average values by the desired multiple of their standard deviations as was demonstrated for the city model. We record the duration of each state (normal, on-fire, and recovery) and the number of emerging fires within the simulated period. The largest relative error of these variables is just below $2\%$.   

In conclusion, we showed that the CARP model is a novel approach to predict and simulate risks. It is particularly useful for modeling cascading catastrophic events and thus has potential applications for analyzing local and global risks. Local risks were demonstrated using simulated fires in cities. However, the CARP model was also successfully used to model global risks in earlier work \cite{Szymanski2015}. A better understanding of globally networked risks is critical to predicting and mitigating them \cite{Helbing2013}. Most of the world's critical infrastructure forms a complex, interconnected network prone to cascading failures with potentially devastating consequences to global stability \cite{perrow1999}. Quantifying the limits of risk prediction, which are bounded by the amount of data, may inform earlier planning and thus potential mitigation of danger of risks spreading and its adverse consequences.

\section*{Methods}
\subsection*{Discrete Model}
Based on the structure of the fire-propagation model, we can simulate the fire cascades throughout the entire network using CARP. At time $t$, each house is either in state $-1$ (recovery), state $0$ (fully operational) or state $1$ (on-fire). Houses in the recovery state are under reconstruction and are immune to fire. Hence only fully operational houses are susceptible to fire. The burning (on-fire) state with certain probability switches to a state of recovery. Each house alternates between these three states. 

The state transition is invoked by four types of Poisson processes. A house $i$ transits from state $0$ to $1$ (on-fire) for internal reasons according to a Poisson process with the intensity $\lambda^{\rm int}_i$. In this event, a fire starts inside of the house, caused by events such as overheating of electrical appliances, unattended stoves or other possible accidents. This house can make this transition if its neighbor $j$ ignites the fire externally through a Poisson process with the intensity $\lambda^{\rm ext}_{ji}$. A transition from state $1$ (on-fire) to $-1$ (recovery) also follows a Poisson process, with the intensity $\lambda^{\rm etg}_i$ (extinguishing the fire). Finally, transition from state $-1$ to $0$ is caused by a Poisson process with intensity $\lambda^{\rm rec}_i$ (completing the recovery process). 

For all the events discussed above the exact time of occurrence is not known, hence we use a discrete time step to accommodate this uncertainty and round up the event time to the nearest integer step. Hence, the evolution of the system can be viewed as a discrete-time series of stochastic processes with three states. For convenience, we assume here that each time unit represents one day of the real world. Consequently, all fires starting on the same day are considered to be starting simultaneously. As shown in Ref. \cite{Szymanski2015} for real time $t$ measured in finite time units (days here) for events generated by a Poisson process, an event happening in at most $\left\lceil t \right\rceil$ time units is identical for the corresponding Bernoulli process. The states of all houses at time $t$ can be represented by a state vector $\vec S(t)$.       

Here, like in Ref. \cite{Szymanski2015}, we assume that experts provide assessments of each house's fire resistance. The value of $N_{1,i}$ represents the likelihood of house i to catch fire internally or externally, and $N_{2,i}$ represents the likelihood that the house fire is extinguished and house is rebuilt over large period of time. In the following, we define control parameters for state transitions with details listed in Table~\ref{tab:Poisson intensity}. The internal risk materialization is controlled by parameter $\alpha$. The external risk materialization is controlled by parameter $\beta$. However, we assume that only the state transition from fully operational to on-fire state is observable, without knowing the actual reason for it. So, impact of an individual parameter $\alpha$ or $\beta$ is hidden from direct observation. In contrast, parameters $\gamma$, controlling recovery and $\delta$ controlling fire extinguishing are independent of each other, so impact of each of the two can be observed in the changes in the evolution of corresponding underlying process. 

We list events and parameters in the order compatible to the way they were ordered in the World Economic Forum model \cite{Szymanski2015} in which only three events (int, ext, rec) and parameters  ($\alpha, \beta, \gamma$) exist, as defined in Table~\ref{tab:Poisson intensity}. We set $N_{1,i} = 0.4$ for large, $0.3$ for medium and $0.2$ for small houses and $N_{2,i} = 0.2$ for large, $0.3$ for medium houses and $0.4$ for small houses. We assign target values to our parameters $\alpha = 0.08$, $\beta = 0.012$, $\gamma = 0.016$ and $\delta = 0.032$.

To summarize, the dynamics progress in discrete steps $t=1,\ldots, T$ and the probability of transition in each step is defined by the intensity of the corresponding Poisson process as shown in Table~\ref{tab:Poisson intensity}. The dynamics described above and shown in Fig. \ref{fig:figure_1}(b) imply that the state of the system at time $t$ depends only on its state at time $t-1$, and therefore the evolution of the state vector $\vec S(t)$ is Markovian. The definitions of parameters and equations mapping them into intensities of Poisson processes are listed in Table~\ref{tab:Poisson intensity}.

\begin{table}[ht]
\begin{center}
\begin{tabular}{|c|c|}
\hline
Name & Definition of variables \\
\hline
$N_{1,i}$ & Likelihood of house $i$ to catch fire (internally or externally) \\
\hline
$N_{2,i}$ & Likelihood of fire to be extinguished and reconstruction started of house $i$ \\
\hline
$\alpha$ & Control parameter for the internal fire materialization process \\
\hline
$\beta$ & Control parameter for the external fire materialization process \\
\hline
$\gamma$ & Control parameter for the recovery process \\
\hline
$\delta$ & Control parameter for the fire extinguished process \\
\hline
$\lambda_i^{int}$ & Intensity for the process of starting fire internally in house $i$: $\lambda_i^{int} = -\alpha \ln N_{1,i}$ \\
\hline
$\lambda_{ji}^{ext}$ & Intensity for the process of externally transferring fire from house $j$ to house $i$: $\lambda_{ji}^{ext} = -\beta \ln N_{1,i}$ \\
\hline
$\lambda_i^{rec}$ & Intensity for the process of completing recovery of house $i$: $\lambda_i^{rec} = -\gamma \ln N_{2,i}$ \\
\hline
$\lambda_i^{etg}$ & Intensity for the process of extinguishing fire in house $i$: $\lambda_i^{etg} = -\delta \ln N_{2,i}$ \\
\hline
$p_i^{int}$ & Probability of internal fire ignition in house $i$: $p_i^{int} = 1 - e^{-\lambda_i^{int}} = 1 - N_{1,i}^{\alpha}$ \\
\hline
$p_{ji}^{ext}$ & Probability of external fire transfer from house $j$ to house $i$: $p_{ji}^{ext} = 1 - e^{-\lambda_{ji}^{ext}} = 1 - N_{1,i}^{\beta}$ \\
\hline
$p_i^{rec}$ & Probability of recovery of house $i$: $p_i^{rec} = 1 - e^{-\lambda_i^{rec}} = 1 - N_{2,i}^{\gamma}$ \\
\hline
$p_i^{etg}$ & Probability of extinguishing fire in house $i$: $p_i^{etg} = 1 - e^{-\lambda_i^{etg}} = 1 - N_{2,i}^{\delta}$ \\
\hline
\end{tabular}

\begin{tabular}{|m{2.5cm}|m{1.1cm}|m{1.1cm}|m{1.1cm}|m{1.1cm}|m{1.1cm}|m{1.1cm}|m{1.1cm}|m{1.1cm}|}
\hline
House type & $\lambda_i^{int}$ & $\lambda_{ji}^{ext}$ & $\lambda_i^{rec}$ & $\lambda_i^{etg}$ & $p_i^{int}$ & $p_{ji}^{ext}$ & $p_i^{rec}$ & $p_i^{etg}$ \\
\hline
Large & 0.0073 & 0.0109 & 0.0257 & 0.0515 & 0.00730 & 0.01094 & 0.02542 & 0.05020 \\
\hline
Medium & 0.0096 & 0.0144 & 0.0192 & 0.0385 & 0.00959 & 0.01434 & 0.01908 & 0.03779 \\
\hline
Small & 0.0129 & 0.0193 & 0.0146 & 0.0293 & 0.01279 & 0.01913 & 0.01455 & 0.02889 \\
\hline
\end{tabular}
\end{center}
\vspace{-0.2in}
\caption{\label{tab:Poisson intensity} Definition of variables, intensities of Poisson processes, and the probabilities of the corresponding Bernoulli processes for the parameter values: $(N_{1,i};\ N_{2,i}) = (0.4;\ 0.2)$ for large, $(0.3;\ 0.3)$ for medium, and $(0.2;\ 0.4)$ for small houses. $\alpha$ = $0.08$, $\beta$ = $0.012$, $\gamma$ = $0.016$ and $\delta$ = $0.032$.}
\end{table} 

With the assumed expert assessments and the created ground truth parameter values for the model, we simulate the evolution of house states for a particular period and record the frequency of emerging fires. In Fig. \ref{fig:figure_1}(d), we show the fraction of time each house is on-fire during one million time steps. The range of this on-fire fraction varies from $0.12$ to $0.32$. The fraction increases when the size or degree of the house increases, but areas of higher density housing suffer higher on-fire fractions than indicated by their degree.

\subsection*{Precision Limit of Maximum Likelihood Estimation (MLE)}
In our approach, we use Maximum Likelihood Estimation (MLE) to recover parameters from ground truth data. The main reason for this choice is that state transitions are governed by independent inhomogeneous probability distributions for which MLE delivers consistency and asymptotic normality with sufficient amounts of observed data \cite{DeGroot2010}. The historical data represents the combined effects of four Bernoulli processes. Our purpose is to recover the unknown parameters $\alpha, \beta, \gamma$ and $\delta$ mapping the expert assessments into event probabilities for Bernoulli processes of our model. We denote $\hat{\alpha}, \hat{\beta}, \hat{\gamma}$ and $\hat{\delta}$ to be the recovered (estimated) values of each parameter.

The use of MLE to find the values of hidden parameters from observed events has not been studied, yet Ref. \cite{DeGroot2010} indicates that it is feasible. In our approach we split one of the parameters of directly observable events into a pair of hidden (and tangled) parameters of two processes and recover these two parameters from observed events.

We denote the unknown parameters as $\theta$. Given the $n$ observations $x_1, x_2, ... x_n$, the likelihood function of this set of observations is defined as:
\begin{equation}
\mathcal{L}\left(\theta\right) = f\left(x_1, x_2, ..., x_n|\theta\right)
\end{equation}
\noindent
When the distribution is discrete, $f$ is a frequency function, defined explicitly by Eq. \ref{likelihood}, and the likelihood function $L(\theta)$ shows the probability of observing the given data. The Maximum Likelihood method \cite{Dempster1977} finds the values of parameters that yield the maximal probability of observing the given data. Logarithms are monotonic and therefore the likelihood and its logarithm have maximum at the same argument. Since the observed data comes from independent distributions, the logarithm of likelihood function can be written as:
\begin{equation}
\ln\mathcal{L}\left(\theta\right) = \sum_{i = 1}^{n} \ln\left(f\left(x_i|\theta\right)\right)
\end{equation} 
\noindent
For continuous and smooth likelihood functions, which is the case here, we can scan the parameter space in order to find the maximum point for $\ln\mathcal{L}(\theta)$. The historical data size is limited, so we need to study how this limitation affects the precision of results. In this section, we derive an estimation of the optimal parameters for large sample sizes. Under appropriate smoothness conditions, the estimate is consistent with large data sets and obeys the asymptotic normality. Detailed description of these conditions can be found in Ref. \cite{Cramer1946}. These conditions can be summarized as: the first three derivatives of the $\frac{\partial \log(f(x_i|\theta))}{\partial \theta}$ are continuous and finite for all values of $x_i$ and $\theta$; the expectation of the first two derivatives of $\frac{\partial \log(f(x_i|\theta))}{\partial \theta}$ can be obtained and the following integral is finite and positive:
\begin{equation}
\int_{- \infty }^{\infty} \left(\frac{\partial \log f}{\partial \theta}\right)^2 f dx
\end{equation}  
\noindent
Let $\theta_0$ denote the true value of the parameter $\theta$ and $\hat{\theta}$ to be its recovered value. Once the smoothness conditions are met, the recovered value $\hat{\theta}$ converges to the true value $\theta_0$ as $n \to \infty$. If we normalize $\hat{\theta}$, we obtain an approximation from a normal distribution:
\begin{equation}
\left(\exists \sigma^2_{MLE} > 0\right): \lim_{n\to\infty}\sqrt{n}\left(\hat{\theta} - \theta_0\right) = N\left(0, \sigma^2_{MLE}\right) 
\end{equation}  
\noindent
Given the definition of Fisher information \cite{Fisher} ($I(\theta)$):
\begin{equation}
I\left(\theta\right) = E \left[\frac{\partial}{\partial \theta} \log f\left(x|\theta\right)\right]^2
\end{equation}
\noindent
The asymptotic normality of MLE can be written as:
\begin{equation}
\lim_{n\to\infty}\sqrt{n}\left(\hat{\theta} - \theta_0\right) = N\left(0, \frac{1}{I\left(\theta_0\right)}\right)
\end{equation}
\noindent
The variance of the normalized estimate decreases as $I\left(\theta_0\right)$ increases. This asymptotic variance to some extent measures the quality of MLE. Although it is hard to compute the variance analytically in our model, we know there exists an estimation limit and the performance of MLE becomes better as the volume of given data increases. In the next section, we demonstrate that the variance indeed decreases as the size of the training dataset increases and the mean values of $\hat{\theta}$ approaches $\theta_0$ with the error approaching $0$.

\subsection*{Parameter Estimation in Fire-propagation Model}
Given the historical data about each house transition in a finite period, we use Maximum Likelihood \cite{Dempster1977} to find values of model parameters that make the model optimally match historical data. State transitions are independent functions with unknown parameter values. Since the historical data is generated from the discrete stochastic process, the likelihood of observing a particular sequence of risk materializations can be written as:
\begin{equation}
\mathcal{L}\left(\vec{S}(1),\vec{S}(2) \cdots, \vec{S}(T)\right) \equiv \prod\limits_{t=2}^{T}\prod\limits_{i=1}^{N} 
P_i(t)^{S_i(t-1)\rightarrow S_i(t)}
\label{likelihood}
\end{equation} 
\noindent
where $T$ is the number of time steps, $N$ is the number of houses in the model, $S_i(t)$ is the current state of house $i$ at time $t$, and $P_i(t)^{S_i(t-1)\rightarrow S_i(t)}$ is the state transition probability of house $i$ from time $t-1$ to time $t$. $\vec{S}(t)$ is the vector of states for each house in the network at time $t$. The logarithm of this likelihood is:
\begin{equation}
\ln\mathcal{L}\left(\vec{S}(1), \vec{S}(2) \cdots, \vec{S}(T)\right) \equiv \sum\limits_{t=2}^{T}\sum\limits_{i=1}^{N}\ln \left( P_i(t)^{S_i(t-1)\rightarrow S_i(t)}\right)
\label{log-likelihood}
\end{equation}
\noindent
In the training process, we compute the probability of state transitions using $p_i^{int}$ and $p_{ji}^{ext}$ for transition into fire, $p_i^{etg}$ for transition into recovery and $p_i^{rec}$ for transition into the fully operational state. Correspondingly, there are three cases of state transitions in the historical data. 

A transition from the fully operational state to the on-fire state ($0 \rightarrow 1$) happens when a house catches fire due to internal or external reasons. The probability of internal ignition is $p_i^{int}$. The probability of external ignition is computed as the complement of the probability that none of the neighbors ignited house $i$. On-fire neighbor $j$ fails to ignite house $i$ with probability $1 - p_{ji}^{ext}$. The product of such individual probabilities over all on-fire neighbors defines the probability that house $i$ is not ignited by external fire: 
\begin{equation}
prod_i^{0 \rightarrow 0} = \prod\limits_{j \in A_i}\left(1 - p_{ji}^{ext}\right)
\end{equation}  
\noindent
where $A_i$ is the set of all on-fire neighbors of house $i$. The complement of this product defines the probability that at least one neighbor ignites house $i$. Adding the probability of such external ignition to the probability of internal ignition, we obtain the total probability of house $i$ catching fire:
\begin{equation}
{P}_i^{0 \rightarrow 1} = p_i^{int} + \left(1 - p_i^{int}\right)\left(1 - \prod\limits_{j \in A_i}\left(1 - p_{ji}^{ext}\right)\right)
\end{equation}
\noindent
Since internal and external ignitions are mutually exclusive, we include a factor of $(1 - p_i^{int})$ in the external ignition probability. Accordingly, the probability of not catching on fire is:
\begin{equation}
{P}_i^{0 \rightarrow 0} = \left(1 - p_i^{int}\right)\prod\limits_{j \in A_i}\left(1 - p_{ji}^{ext}\right)
\end{equation}
\noindent
A transition from being on fire to recovery state happens when the fire is extinguished and rebuilding process starts. This probability is defined as:  
\begin{equation}
{P}_i^{1 \rightarrow -1} = p_i^{etg}
\end{equation}
\noindent
Transition from recovering to fully operational state ($-1 \rightarrow 0$) happens when a house is completely rebuilt and becomes fully operational. The corresponding probability is: 
\begin{equation}
{P}_i^{-1 \rightarrow 0} = p_i^{rec}
\end{equation}
\noindent
The maximum likelihood parameters are obtained by summing the logarithms of corresponding probabilities. After scanning the potential ranges of the model parameters, we find the globally optimal values that maximize the likelihood of the historical data. The closeness of the recovered parameters to their values set in the simulations measures how precisely our model captures the dynamics of the system.

\section*{Acknowledgments}
This work was partially supported by DTRA Award No.HDTRA1-09-1-0049, and by the Army Research Laboratory under Cooperative Agreement Number W911NF-09-2-0053 (the ARL Network Science Collaborative Technology Alliance). The views and conclusions contained in this document are those of the authors and should not be interpreted as representing the official policies either expressed or implied of the U.S. Army Research Laboratory or the U.S. Government. 

\section*{Author Contributions}
BKS conceived the research;
XL, AM, GK, BKS, and JB designed the research;
XL, AM, BKS implemented and performed numerical experiments and simulations; 
XL, AM, GK, BKS, and JB analyzed data and discussed results;
XL, AM, GK, BKS, and JB wrote and reviewed the manuscript.

\section*{Competing Financial Interests Statement}
The authors declare no competing financial interests.

\end{document}